\documentclass[showpacs,showkeys,pre]{revtex4}
\usepackage{graphicx}
\usepackage{dcolumn}
\usepackage{amsmath}
\usepackage{latexsym}
\usepackage{longtable}

\begin {document}

\title
{Temperature dependence of phenanthrene cavity radius in apolar
solvents and in water}
\author
{ I. A. Ar'ev$^1$ and N. I.  Lebovka$^{2}$} \affiliation { $^1$
Institute of Colloid and Water Chemistry named after A. V.
Dumanskii, NASU, bulv. Vernadskogo, 42, 03142,Kyiv, Ukraine\\ $^2$
Biocolloid Chemistry Institute named after F. D. Ovcharenko, NASU,
bulv. Vernadskogo, 42, 03142,Kyiv, Ukraine}
\begin{abstract}
Phenanthrene, a three ring aromatic hydrocarbon, is used as a
model substance for study of intermolecular interactions in dilute
solutions of organic solvents and water by spectroscopic method.
Temperature dependencies of shift of the S$_2\leftarrow$S$_0$
spectrum of phenanthrene dissolved in apolar solvents and in
liquid water are studied in this work. The spectroscopic data are
used for analysis of the cavity radius $R$ for phenanthrene
molecule in solvents. It is shown that the value of $R$ increases
with temperature in organic solvents. In contrast, for water
solution, the value of $R$ initially grows with temperature
increase (in the interval of 273.3-296 K), but becomes constant at
higher temperatures. The changes of water structure in the
neighbourhood of a phenanthrene molecule and probable hypothesis
about causes of $R$ value constancy in high-temperature region are
discussed. Conclusion about strengthening of the water structure
in an aqueous cage with phenanthrene molecule inside is done.

\end{abstract}

\maketitle

\section{INTRODUCTION}
\label{INTRO}

 Solute-solvent distribution function is the main
characteristic for description of the structure of different
molecular solutions at molecular level. A rather popular problem
in this field is one related to study of the different aspects of
hydrophobic hydration. Hydrophobic effects control solubility of
apolar molecules in an aqueous solvents, formation of micelles,
protein folding, etc. \cite{1}. Hydrophobic hydration is a driving
force of the hydrophobic effect. Unusual distribution of water
molecules around an apolar solute or an apolar fragment of
biphyllic molecule is the unique property of hydrophobic
hydration. In spite of sustained efforts undertaken during last
decades and existence of a lot of theoretical speculations, this
matter is still poorly studied experimentally.

This situation is caused by the lack of effective empirical tools
for structural studies in the nearest vicinity of hydrophobic
molecules dissolved in water. The aqueous solubility of
hydrophobic molecules in the majority of cases are very low.
Traditional experimental methods, e.g. neutrons or x-rays
scattering, are ineffective in application to such diluted
solutions.

The same difficulty is encountered also for nonaqueos diluted
solutions, where both solute and solvent molecules are apolar.
Solutions of apolar compounds in apolar solvents can serve as
reference systems. For these solutions the contribution of
electrostatic interactions can be neglected.

The spectral shift method \cite{2} can be effective in such a
situation, though it is somewhat rough. As a rule, transfers of a
molecule from one environment to another one induces shifts in its
electronic spectrum. The spectral shift $\Delta\nu$ depends on
molecular environment of the test molecule. The method of spectral
shifts can be used for estimation of the mean distance between the
solute and solvent molecules when both solute and solvent
molecules are apolar. This distance may be identified with a
certain approximation as $R +r$, where $R$ is the radius of the
cavity, which contains a solute molecule, and $r$ is the radius of
the solvent molecule. Note that the first maximum of the
solute-solvent pair distribution function is approximately
localized at $R+r$. This method can be applied also for polar
solvents, though with definite restriction: the permanent electric
fields induced on the solute by solvent molecules should be
compensated.

Here we present results of investigation of the shift dependencies
in S$_2\leftarrow$S$_0$ spectrum $\Delta\nu$ of phenanthrene
dissolved in some apolar solvents, in methanol and in water in the
wide temperature ranges. Our purpose is to examine changes of $R$
with temperature. The paper is organized as follows. In Sec.
\ref{SPECTRAL} we give introductory information about the
solvent-induced spectral shift method. In Sec. \ref{EXPERIMENTAL}
the experimental procedures are described. Experimental results
are presented and discussed in Sec. \ref{RESULTS}. Conclusions are
summarized in Sec. \ref{CONCLUSIONS}

\section{THE SPECTRAL SHIFT METHOD}
\label{SPECTRAL}

Let an apolar molecule be transferred from a rarified gas into the liquid
solvent. Then, in approximation of the dielectric continuum, the spectral
shift is

\begin{equation}
\Delta \nu = - C\phi (R) f(n) +\Delta \nu _{o} \label{Eq1}
\end{equation}
\noindent where, $C$ is a positive factor that characterizes the
solute molecule and $f(n)= (n^2 - 1)/(n^2 + 2)$ is the
Lorentz-Lorenz function, $n$ is the refractive index of the
solvent, $\Delta \nu _{o}$ is related to contributions other than
dispersion interactions. The $\phi (R)$ is a function of the
cavity radius. When the solute molecule is located in the cavity
center, $\phi(R) = 1/R^3$. When the solute molecule is in contact
with the cavity surface, $\phi(R) = R^3/(r_{u}^{3}(2R -
r_{u})^{3})$, where $r_{u}$ is the solute radius.

Eq. \ref{Eq1} is good for approximation only in a case when the
solute oscillator strength lies in the interval from 0.1 to 1 (see
\cite{2} and references cited therein).

Eq. \ref{Eq1} can be rewritten as
\begin{equation}
\phi(R)=(\Delta \nu_o -\Delta\nu )/(C f(n)). \label{Eq2}
\end{equation}

This equation may be used for determination of the temperature
dependence of $R(T)$. In particular, in a case when the solute
molecule is situated in the cavity center, it follows from Eq.
\ref{Eq2} that

\begin{equation}
\frac{{R\left( {T} \right)}}{{R\left( {T_{ref}}  \right)}} =
\left[ {\frac{{\left( {\Delta \nu _{0,ref} - \Delta \nu _{ref}}
\right)f\left( {n} \right)}}{{\left( {\Delta \nu _{0} - \Delta
\nu}  \right)f\left( {n_{ref}}  \right)}}}
\right]^{1/3},\label{Eq3}\end{equation}

\noindent where subscript \textit{ref} shows that the property was
measured at some reference temperature $T_{ref}$.

The temperature coefficient of the spectral shift may be defined as

\begin{equation}
\alpha _{s} = (1/\Delta \nu )(d\Delta \nu /dT). \label{Eq4}
\end{equation}

Then, in the case when $\Delta \nu _{o}$ contribution to Eq.
\ref{Eq1} may be neglected (see \cite{2} for more detailed
discussion), and the solute is located in the cavity center, we
obtain

\begin{equation}
\alpha _{s} = -\alpha _{v} - \alpha _{c},\label{Eq5}
\end{equation}
\noindent where $\alpha _{v}$ is the volume expansion temperature
coefficient of the bulk solvent, and $\alpha _{ñ} = (3/R)(dR/dT)$
is the volume expansion temperature coefficient of the cavity,
occupied by a solute. This equation determines the lower limit of
$\alpha_{s}$.

\section{EXPERIMENTAL}
\label{EXPERIMENTAL}

All absorption spectra were recorded by spectrophotometer Specord UV Vis
using the following procedure. First, the reference spectrum of lanthanide
salts dissolved in fused quartz was recorded, then the spectrum of the
studied sample was recorded and then the reference spectrum was recorded
again. The registration scale was 20.2 cm$^{-1}$/mm, the registration rate
was 500 cm$^{-1}$/min. The spectrograms were suitable for the next treatment
if the distance between the band maximums in the first and the last
reference spectrograms did not exceed 2 cm$^{-1}$. The wave number of the
phenanthrene band maximum was measured for each spectrogram and averaged for
all the data. The root-mean-square errors (rms) of wave-number positions did
not exceed 2 cm$^{-1}$.

We used the twice-distilled water, chemically pure Phenanthrene and pure
organic solvents for chromatography. The concentration of phenanthrene was
of the order of 10$^{-6}$ mol/dm$^{3}$ for aqueous solution (5 cm cell) and
of the order of 10$^{-4}$ - 10$^{-5}$ mol/dm$^{3}$ (1 - 4 mm cells) for
other solutions. Cylindrical cells and specially designed temperature
chamber were used. Water was serving as a heat-carrier. Temperature of
solutions was stabilized with a precision of 0.1 K.

Temperature shifts of phenanthrene vapor spectrum were measured using a
procedure, different from that applied for solutions. The pressure of
phenanthrene vapor in the preliminary pumped out and soldered 2-cm quartz
cell was of the order of 10$^{-2}$ mm Hg. The sample was heated in the
special chamber to the temperatures from 360 to 615 K. The temperature
variations did not exceed 0.2 K.

The temperature dependence of the long-wave band in $S_{2}
\leftarrow S_{0}$ Phenanthrene vapor spectrum was approximated by
the linear equation

\begin{equation}
\nu =\nu _{o}-k_\nu T,\label{Eq6}
\end{equation}
\noindent where $\nu _{o}=35654 \pm 18$ ñm$^{-1}$, $k_\nu =1.175
\pm 0.038$ ñm$^{-1}$K$^{-1}$, $Ò$ is the absolute temperature. The
correlation coefficient of rms approximation was 0.999. Eq.
\ref{Eq6} is qualitatively in agreement with analysis data for the
$S_{1} \leftarrow S_{0}$ temperature shift of anthracene
transition in the gas phase \cite{3}. So, we can conclude that
this transition is of the same nature as $S_{2} \leftarrow S_{0}$
transition of phenanthrene \cite{4}.

All necessary reference data were taken from \cite{5,6,7,8,9}.

\section{RESULTS AND DISCUSSION}
\label{RESULTS}

\begin{figure}
\begin{center}
\includegraphics[width=12.0cm]{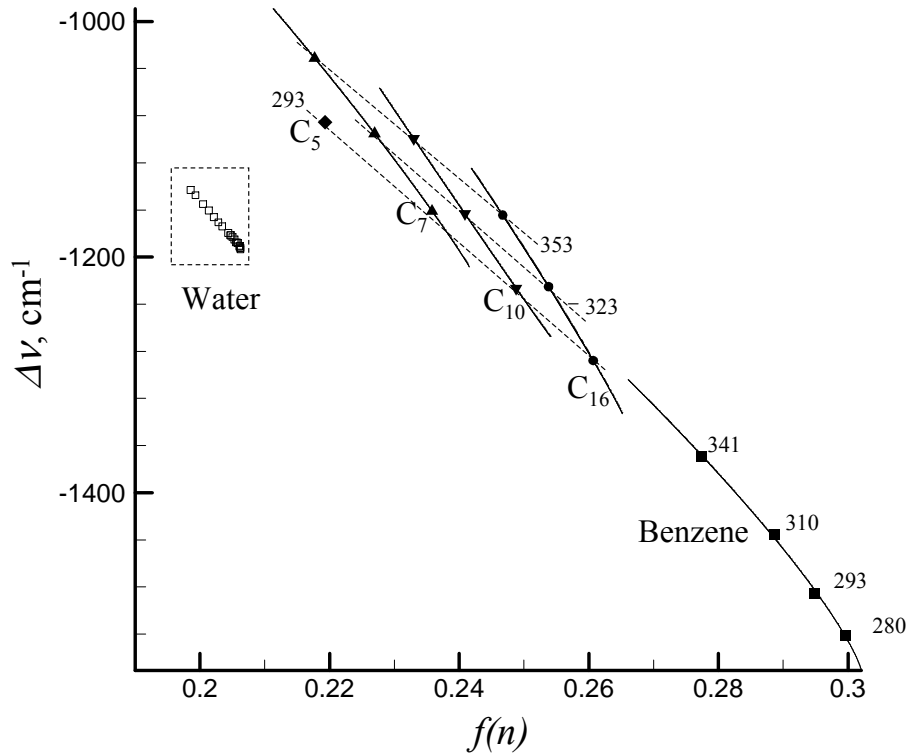}
\end{center}
\caption{The $S_{2} \leftarrow S_{0}$ spectral shift, $\Delta \nu
$, of phenanthrene versus Lorentz--Lorenz function, $f(n)= (n^2 -
1)/(n^2 + 2)$, in dilute solutions of n-pentane (C5), n-heptane
(C7), n-decane (C10), n-hexadecane (C16), benzene, and  water. The
temperature T in K is indicated near the experimental points. The
more large scale presentation for water is given at Fig.
\ref{f02}.} \label{f01}
\end{figure}

Figure \ref{f01} shows experimental data for the spectral shifts
$\Delta \nu $ of phenanthrene versus Lorentz--Lorenz function
$f(n)= (n^2 - 1)/(n^2 + 2)$ in dilute solutions of organic
solvents , and water at different temperatures.

Table 1 shows values of $\alpha _{v}$, $\alpha_{s}$, $-\alpha
_{v}-\alpha_{s}$, for phenanthrene solutions at $T$=293 K. Data
for anthracene solutions are included for comparison (were taken
from \cite{10} or re-measured).

Although the anthracene polarizability exceeds that of
phenanthrene, the observed values of $\alpha_{s}$ and $\alpha
_{c}=-\alpha_{v}-\alpha_{s}$ are of the same order of magnitude
within the experimental errors. It is surprising at first glance,
because the more is polarizability, the more is intermolecular
attraction interactions. So, values of $\alpha_{s}$ and $\alpha
_{c}$ for these substances should be noticeably different. But
phenanthrene molecule is more compact than anthracene one. So the
cavity size for this molecule is more small and the attraction
interactions for the phenanthrene are more pronounced than for the
anthracene.

\begin{widetext}

Table 1. Solvent thermal expansion coefficients ($\alpha _{v}$),
shift temperature coefficients ($\alpha _{s}$), and -$\alpha
_{v}$-$\alpha _{s}$ at 293 K for phenanthrene (Ph) and anthracene
(A). The temperature ranges used for determination of $\alpha
_{s}$ values are shown in parentheses.

\begin{longtable}
{|p{70pt}|p{55pt}|p{106pt}|p{106pt}|p{40pt}|p{40pt}|} a & a & a &
a & a & a  \kill \hline \raisebox{-1.50ex}[0cm][0cm]{Solvent}&
\raisebox{-1.50ex}[0cm][0cm]{$\alpha _{v}$, 10$^{-3}$, \newline
K$^{-1}$}& \multicolumn{2}{|p{219pt}|}{$\alpha _{s}$,
10$^{-3}$K$^{-1}$} & \multicolumn{2}{|p{92pt}|}{-$\alpha _{s}
$-$\alpha _{v}$, 10$^{-3}$K$^{-1}$}  \\ \cline{3-6}
 &
 &
Ph& A& Ph& A \\ \hline Benzene& 1.20& -1.69 (279.9-341.1 K)& -1.70
(283-333 K)& 0.49$^{}$& 0.50 \\ \hline n-Heptane& 1.13& -1,94
(277.6- 360.1 K)& -1.92 (274.2-309.1 K)& 0.81$^{}$& 0.79 \\ \hline
n-Decane& 1.05& -1.75 (282.4-338.2 K)& -1.75 (275.4-353.1 K)&
0.70$^{}$& 0.70 \\ \hline n-Hexadecane& 0.88& -1.49 (293.1 - 355.6
K)& -& 0.61& - \\ \hline \textit{}Methanol& 1.14& -1.63
(276.4-323.3 K)& -1.64 (293.1 - 324.3 K)& 0.48& 0.50 \\ \hline
Acetone& 1.41& -& -2.25 (276.6-321.7, K)& -& 0.84 \\ \hline
\end{longtable}

\end{widetext}

The values of $-\alpha_{s} - \alpha _{v}$ represent temperature
expansion of cavities in apolar solvents. These values for polar
solvents are also non-zero and positive.

\begin{figure}
\begin{center}
\includegraphics[width=12.0cm]{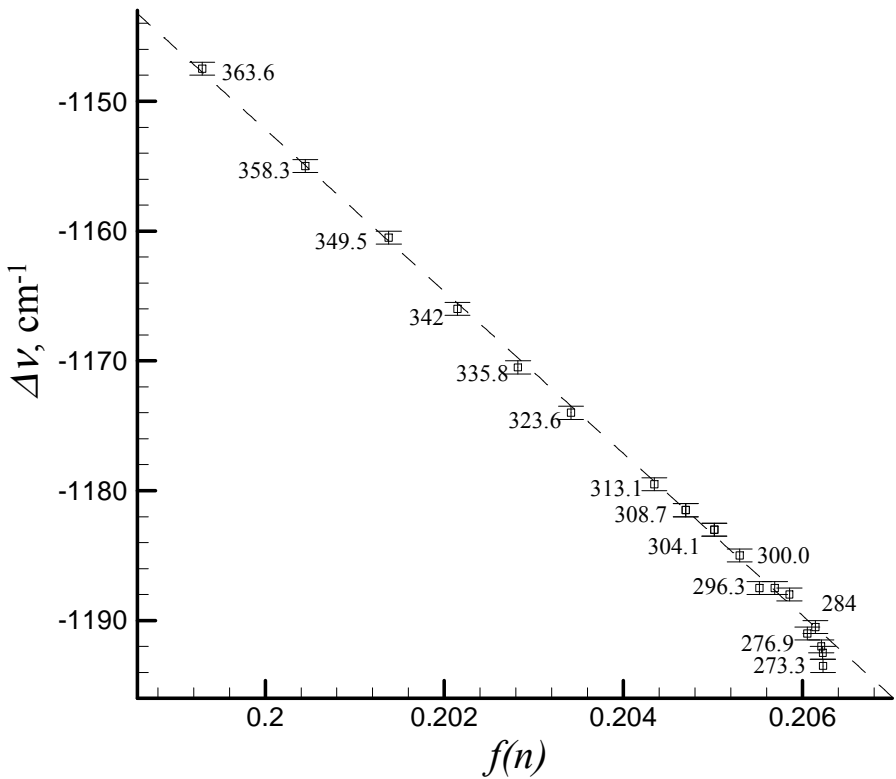}
\end{center}
\caption{The $S_{2} \leftarrow S_{0}$ spectral shift, $\Delta \nu
$, of phenanthrene versus Lorentz--Lorenz function, $f(n)= (n^2 -
1)/(n^2 + 2)$, in dilute water solution. The temperature T in K is
indicated near the experimental points.} \label{f02}
\end{figure}

Data for water spectral shift $\Delta \nu $ in the temperature
range from 296 to 363.6 K the may be approximated by linear
equation \ref{Eq1}. The equation parameters are $C\phi(R) = 6250
\pm 50$ cm$^{-1}$, $\Delta \nu _{0}= 97 \pm 10$ cm$^{-1}$ , and
correlation coefficient is 0.998 (dashed line in Fig. \ref{f02}).
It should be noted that these parameters were estimated proceeding
from an assumption that terms of Eq. \ref{Eq6} are $\nu
_{o}=35654$ ñm$^{-1}$ and $k_\nu =1.175$ ñm$^{-1}$K$^{-1}$
exactly. Taking into account errors of estimations, one cannot
state that $\Delta \nu _{0} \approx 0$ and that means that
electrostatic fields on phenanthrene can be canceled effectively.
At low temperatures (in the interval of 273.3-296 K) the
pronounced deviations from linear relation is observed.

\begin{figure}
\begin{center}
\includegraphics[width=12.0cm]{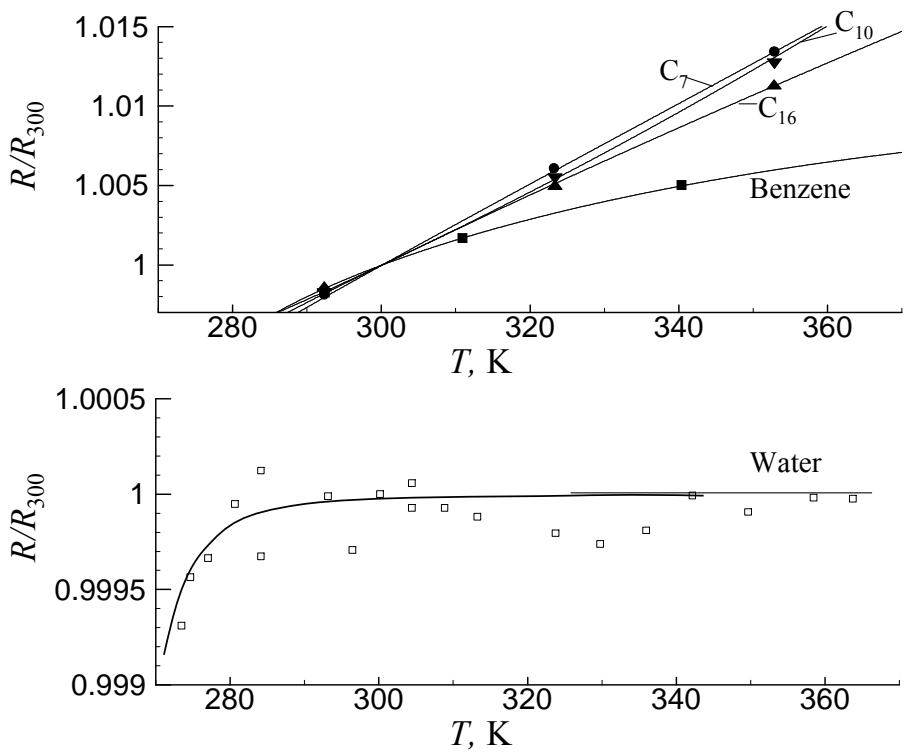}
\end{center}
\caption{Temperature dependencies of relative phenanthrene cavity
radius $R(T)/R_300$ in n-heptane (C$_{7}$), n-decane (C$_{10}$),
n-hexadecane (C$_{16}$), benzene and water.} \label{f03}
\end{figure}

Figure \ref{f03} presents the temperature dependencies of relative
phenanthrene cavity radius $R(T)/R_{300}$ in organic solvents and
water estimated from Eq. \ref{Eq3}. Here ${R}_{300}$ corresponds
to the cavity radius at reference temperature $T=300$ K.

The cavity radius continuously grows with temperature in organic
solvents. It is interesting to note that $R$ value of grows more
noticeably within congeneric series C$_7$ - C$_{16}$ than in
benzene solutions. In contrast the $R$ value of water solution
initially grows with temperature increase (in the range of
273.3-296 K), but becomes constant at higher temperatures. The
explanation of this experimental observation is not trivial.

This fact can evidence higher structurization of the aqueous cage
with phenanthrene molecule inside at high temperatures. It is in
accordance with conclusion made in \cite{11} regarding
strengthening of the hydrogen bonds in the first hydration shell
of a solute molecule and formation of a more organized structure
around the solute. Tanaka \cite{12} has also observed in his
simulations the effect of structural enhancement in water near a
large cavity.

It is possible that at temperatures exceeding 296 K, the H-bonds
between water molecules get strengthened in the nearest vicinity
of phenanthrene to such extent that thermal factor not affect
them. Tomchuk and Krasnoholovets \cite{13} have shown that if the
chain of water molecules is not perturbed thermally, then all the
protons in the chain vibrate coherently together jumping from one
O - H bond to the other one. Two ions of the opposite signs arise
on the ends of such chain as a result of their first jump, the
next jump restores the initial position and so on. This process
has a tunneling nature \cite{13}. Note that the number of O - H
bonds decreases by one during the first jump. Hence, the total
energy of the chain should increase by the energy of one H-bond
breakage. The loss of the bond should be compensated by electronic
polarization that results in the strengthening of the hydrogen
bonds. But the energy of the chain does not increase. The
molecular beams experiments \cite{14} support this assumption. The
electronic polarization of water molecules can also play an
important role in process of electron tunneling through the water
layers \cite{15}. So, the coherent proton vibration can promote
stabilization of the chains of water molecules.

\section{CONCLUSIONS}
\label{CONCLUSIONS}

We have measured temperature shifts of S$_{2}\leftarrow $ S$_{0}$
in the spectra of gaseous phenanthrene and phenanthrene dissolved
in some organic apolar and polar solvents and in water. The
temperature expansions of cavities containing solute molecules
inside, $\alpha _{c}$'s, were calculated for the apolar solvents
as differences between temperature coefficient of the shift,
$\alpha _{s}$, and temperature expansion of the solvent, $\alpha
_{v}$, and were compared with the same values for anthracene
solutions. It was shown that the value of $R$ for phenanthrene
increases with temperature in organic solvents. The aqueous system
shows different properties at temperatures higher than 296 K and
below this temperature. In contrast, the value of $R$ in water
solution of phenanthrene initially grows with temperature increase
(in the range of 273.3-296 K), but becomes constant at higher
temperatures. These phenomena were related to strengthening of the
hydrogen bonds in the first hydration shell of a solute molecule
and formation of a highly organized structure around the solute.

\section*{ACKNOWLEDGEMENTS}
Authors thank Dr. N.S. Pivovarova for her help with preparation of
the manuscript.

\end{document}